\documentclass[traditabstract]{aa}
\usepackage{txfonts}
\usepackage{natbib}
\usepackage{graphicx}
\bibpunct{(}{)}{;}{a}{}{,}

\begin{document}
  \title{Excitation temperature of $C_2$ and broadening of the 6196\,\AA~diffuse interstellar band
  \thanks{Based on data collected at the ESO (8m telescope at Paranal, 3.6m and 2.2m telescopes at La Silla) and observations made with the 1.8m telescope in South Korea and the Cassegrain Fiber Environment in Hawaii.}}

   \authorrunning{M. Ka{\'z}mierczak et al.}
   \titlerunning{Excitation temperature of C2 and broadening of the DIB 6196A}
   \author{M. Ka{\'z}mierczak \inst{1}
          \and P. Gnaci{\'n}ski \inst{2}
   	  \and M. R. Schmidt \inst{3}
   	  \and G. Galazutdinov \inst{4}
   	  \and A. Bondar \inst{5}
   	  \and J. Kre{\l}owski \inst{1}
          }

        \institute{Centre for Astronomy, Nicolaus Copernicus University,
	      Gagarina 11, 87-100 Toru{\'n}, Poland\\
              \email{kazmierczak@astri.uni.torun.pl,jacek@astri.uni.torun.pl}
         \and Institute of Theoretical Physics and Astrophysics,
		University of Gda{\'n}sk, Wita Stwosza 57, 80-952 Gda{\'n}sk, Poland\\
             \email{pg@iftia9.univ.gda.pl}
         \and Nicolaus Copernicus Astronomical Center, 
	      Rabia{\'n}ska 8, 87-100 Toru{\'n}, Poland\\
              \email{schmidt@ncac.torun.pl}
         \and Korea Astronomy and Space Science Institute, ICAP,
		61-1, Hwaam-dong, Yuseong-gu, Daejeon, 305-348, Republic of Korea\\
              \email{gala@boao.re.kr}
         \and International Centre for Astronomical 
 	and Medico-Ecological Research, Terskol, Russia\\
              \email{arctur@inet.ua}\\
		}       
       
\date{Received / Accepted }

%
%
\abstract{This paper presents a finding of the correlation between the width of a strong diffuse interstellar band at 6196\,\AA~and the excitation temperature of $C_2$ based on high resolution and high signal-to-noise ratio spectra. The excitation temperature was determined from absorption lines of the Phillips \mbox{$\rm A^{1}\Pi_u-X^{1}\Sigma^{+}_g$} and Mulliken \mbox{$\rm D^{1}\Sigma^{+}_u-X^{1}\Sigma^{+}_g$} systems. 
The width and shape of the narrow 6196\,\AA~DIB profile apparently depend on the $C_2$ temperature, being broader for higher values.}

\keywords{ISM: molecules and diffuse bands}

\maketitle

\label{firstpage}

%
%

\section{Introduction}

Diffuse interstellar bands (DIBs), which were discovered by Heger in 1922, have remained unidentified since that time in spite of being the subject of much observational and theoretical research. Since the first discovery of the strong features at 5780\,\AA~and 5797\,\AA,~the list of these spectral bands has continued to grow (at present we know 380 DIBs \citep{Hobbs2008}). Their origin is still unknown, but usually they are related to some complex, carbon-bearing molecules \citep{Douglas1977,Fulara2000,Fulara2003}.

Important information allowing identification of diffuse interstellar bands may come from an analysis of their profiles. The first description of the profiles of strong diffuse bands was given by \citet{Herbig1975} and then by \citet{Westerlund1988}. The task is quite difficult because very often we observe more than one interstellar cloud along one line of sight, and Doppler splitting likely modifies the profiles of DIBs \citep{Herbig1982}. Observed DIB profiles may be formed in several clouds, which are characterized by different physical parameters. Different conditions in interstellar clouds may lead to the profile variations from object to object reported by \citet{Galazutdinov2002}. The profiles of diffuse interstellar bands often have substructures; \citet{Sarre1995} showed that the substructures inside the profiles of strong DIBs are very narrow and shallow. 

The diffuse interstellar band at 6196\,\AA~is one of the sharpest DIBs. Its profile sometimes shows a substructure intrinsic to it \citep{Galazutdinov2008} and the width (FWHM) is apparently variable \citep{Galazutdinov2002}. Both of these phenomena make the feature a very interesting object, likely reflecting the physical parameters of intervening clouds.  

The vast majority of the already identified molecules in interstellar clouds are polar species. At the low gas density of diffuse clouds, the rate of cooling through spontaneous emission significantly exceeds collisional heating and the excitation of rotational levels is governed mostly by the microwave background radiation. Therefore we cannot derive information about a cloud's kinetic temperature from polar molecules. This is however possible with homonuclear ones. The long lifetime of excited rotational states results collisions now being frequent enough to thermalize the molecule. $C_2$, the simplest multicarbon molecule, is particularly interesting. $C_2$ is quite easily observed by ground-based instruments in the optical spectrum through (1,0), (2,0), (3,0) and (4,0) bands of the Phillips system. The absorption lines up to the rotational number J''=18 can be identified, allowing derivation of individual column densities and estimation of the excitation temperature. The same parameters can be determined from the Mulliken $\rm D^{1}\Sigma^{+}_u-X^{1}\Sigma^{+}_g$ (0,0) band in the UV range.  

It has been conjectured that long carbon chains may be responsible for some DIBs \citep{Douglas1977,Maier2006}, but some attempts to identify features of any chain longer than $C_3$ have failed \citep{Maier2002,Galazutdinov2002}. There is thus no evidence that long chains cause DIBs. However, it is interesting to search for the relations between the profiles of diffuse interstellar bands, carried potentially by long carbon chains, and their smallest precursors ($C_2$ or $C_3$).

The goal of this paper is to study the relations between the profile widths of the strong diffuse interstellar band at 6196\,\AA~and the excitation temperatures of $C_2$. In the next section we describe the observations. A general discussion and summary of our conclusions are given in sections 3 and 4 respectively.

%
%
\section{Observations}
%
%
\subsection{Program stars}
%
%

We have chosen spectra containing the interstellar molecular bands of $C_2$ and the DIB at 6196\,\AA~towards eleven objects acquired using different instruments. Objects with intermediate colour excess and a single (or one strongly dominant) velocity component were selected (Table 1) to make it likely that analysed features originate from single clouds. The profiles of $KI$, $CH$ and $CH^+$ lines were checked for the existence of more than one dominating Doppler component visible in our high resolution and high signal-to-noise ratio spectra. Almost all of these lines in the spectra of the programme stars are free of any observable Doppler splitting. One exception is the $CH^+$ line towards HD\,204827. Previous papers \citep{Pan2004,Welty2001} show multiple components in very high resolution spectra in the interstellar lines towards HD\,204827 and a very weak Doppler component in KI and CH toward HD\,148184. These authors also show that almost all of the program objects have multiple components in NaI; however such components may be unrelated to molecular ones or to DIBs \citep{Bondar2007}. Moreover, the latest survey of diffuse interstellar bands was based on HD\,204827 spectra \citep{Hobbs2008}. The lack of Doppler splitting in the interstellar $CH$ (4300\,\AA)~line in our spectra is demonstrated in Figure 1 (see also \mbox{Figure 9).} Doppler splitting is easier to detect when the observed feature is narrower. Thus, while the observed CH profiles do not show Doppler components, the latter cannot cause the observed broadening of the much broader DIB. Naturally it is not possible to prove that DIB carriers are spatially correlated to any atomic and molecular species; the only hints come from their similar Doppler shifts in different objects.

\begin{table}
{\vbox{\footnotesize
	\tabcolsep=1.5pt
\caption{Basic data of our program stars, columns 'DIB' and '$C_2$' identify sources of the 6196\,\AA~DIB profile and of the $C_2$ band, 'S/N' determines the signal-to-noise ratio in the range of the spectrum near to a 6196\,\AA~feature.
}
\label{Table 1.}
\centering
\begin{tabular}{lcccccc}
\hline \hline
object & name& Sp/L* & E(B-V)* & DIB & S/N & $C_2$\\
\hline
HD\,23180  & o Per      & B1III	  & 0.27 & GECKO & 600& BOES\\
HD\,110432 &            & B2pe    & 0.48 & FEROS & 300& $^a$\\
HD\,147888 &$\rho$ Oph D& B5V     & 0.47 & FEROS & 500& $^b$\\
HD\,147889 &            & B2V     & 1.07 & HARPS & 500& UVES\\ 
HD\,148184 &$\chi$ Oph  & B2Vn,el & 0.52 & HARPS & 600& UVES\\
HD\,149757 &$\zeta$ Oph & O9.5V   & 0.32 & CES   &1000& $^c$\\
HD\,154368 &            & O9.5Iab & 0.73 & UVES  & 300& UVES\\
HD\,163800 &            & O7IIIf  & 0.60 & HARPS & 500& UVES\\
HD\,169454 &   V430 SCT & B1Ia    & 0.93 & UVES  & 400& UVES\\
HD\,179406 & 20 Aql     & B3V     & 0.33 & HARPS & 600& UVES\\
HD\,204827 &            & B0V     & 1.11 & BOES  & 300& $^d$\\
\hline
\multicolumn{7}{l}{ * Thorburn et al. (2003), Hunter et al. (2006)}\\
\multicolumn{7}{l}{$^a$ Calculated by Sonnentrucker et al. (2007) using the equivalent}\\ 
\multicolumn{7}{l}{widths of van Dishoeck \& Black (1989) (Phillips system)}\\
\multicolumn{7}{l}{$^b$ Sonnentrucker et al (2007) (Mulliken system)}\\
\multicolumn{7}{l}{$^c$ Calculated by Sonnentrucker et al. (2007) using the equivalent}\\ 
\multicolumn{7}{l}{widths of Hobbs \& Campbell (1982) (Phillips system)}\\
\multicolumn{7}{l}{$^d$ Calculated in this work using the equivalent}\\ 
\multicolumn{7}{l}{widths of {\'A}d{\'a}mkovics et al. (2003) (Phillips system)}\\
\end{tabular}
}}
\end{table}

\begin{figure}
\centering
\includegraphics[width=0.35\textwidth]{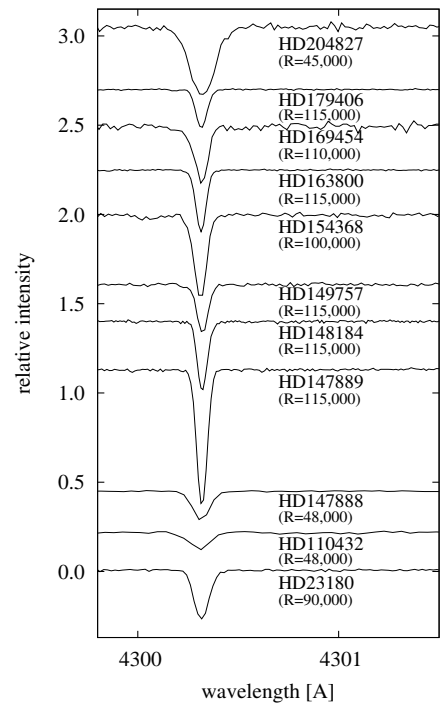}
\caption{The demonstration of a lack of Doppler splitting in the interstellar molecule CH 4300\,\AA~ in our spectra of the programme stars}
\label{fig1}
\end{figure}

%
%
\subsection{Spectra}
%
%

This paper is based on six sets of observed spectra. Strengths of the Phillips (\mbox{$\rm A^{1}\Pi_u-X^{1}\Sigma^{+}_g$}) bands of the $C_2$ molecule were measured in the spectra from UVES \citep{Bagnulo2003} (one exception is HD\,23180 where we used data from BOES), because this spectrograph allows us to observe three vibrational bands of this system. We measured the absorption lines from the three bands (where possible) of $C_2$ (1-0) \mbox{10133 - 10262\,\AA}, (2-0) \mbox{8750 - 8849\,\AA} and (3-0) \mbox{7714 - 7793\,\AA} to make our estimates of the excitation temperature more reliable. For HD\,110432, HD\,147888, HD\,149757 and HD\,204827 the excitation temperature of $C_2$ was taken from previous papers (from the Phillips system as well as the Mulliken band; see Table 1).
The diffuse interstellar band at 6196\,\AA~was analysed using the spectra from GECKO, FEROS, HARPS, CES, UVES and BOES depending on which one was available or had a better signal-to-noise ratio in this range of the spectrum, as shown in \mbox{Table 1.} 

\begin{figure}
\centering
\includegraphics[width=0.45\textwidth]{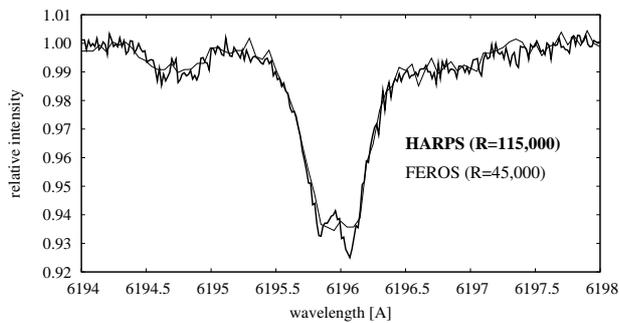}
\caption{
The comparison of the profile width of the 6196\,\AA~DIB in the spectra made with different spectral resolution towards HD\,147889.
}
\label{fig2}
\end{figure}

One of the set of spectra used to analyse the interstellar absorption features was acquired with the spectrograph HARPS. High Accuracy Radial velocity Planet Searcher (HARPS) at the ESO La Silla 3.6m telescope is dedicated to the discovery of extra-solar planets and it can measure radial velocities with the highest accuracy currently available. HARPS produces the spectra with a resolving power of 115,000 in the spectral range  3780-6910\,\AA, with a signal-to-noise ratio of 500.

UV-Visual Echelle Spectrograph (UVES) is the high-resolution spectrograph of the VLT fed by the Kueyen telescope of the ESO Paranal Observatory, Chile. The high quality of the UVES spectrograph allows us to obtain excellent spectra (with resolution 80,000 - 110,000 and signal-to-noise ratio above 200) over a wide range of the spectrum (3000 - 11,000\,\AA).

BOES is the echelle spectrograph (R=\, 30,000; 45,000; 90,000) attached to 1.8m telescope of the Bohyunsan Optical Astronomical Observatory, in South Korea. The working wavelength range is from
3500\AA to 10000\AA.

GECKO was an echelle coud{\'e} spectrograph of the the Canada-France-Hawaii Telescope. Its spectral resolution can reach up to 120,000.

The Fiber-fed Extended Range Optical Spectrograph (FEROS) is a high-resolution (R=48,000) astronomical echelle spectrograph with very high efficiency and a wide wavelength range. It is operated at the 2.2m telescope of the European Southern Observatory (ESO) in La Silla, Chile.

CES was the ESO Coud{\'e} Echelle Spectrograph fed by the 3.6m telescope, at La Silla Observatory, Chile. Observations were made with the highest resolving power, R=220,000, using the Very Long Camera.

The spectra used to analyse the profiles of the 6196\,\AA~feature were obtained with different resolution, but this does not change the widths of the reasonably broad DIB profiles. Figure 2 compares the 6196 DIB profile width in the spectra of HD\,147889, with the resolutions of 115,000 (HARPS) and 45,000 (FEROS); the FWHM estimated from these two instruments is the same (0.52\,\AA), but in the spectra with low resolution the substructures inside the profiles are hard to trace.

%
%
\subsection{Data reduction}
%
%

Reduction of the spectra was made using the Dech20T code \citep{Galazutdinov1992}. This program allows location of the continuum, measurements of the line equivalent widths, a full width at half maximum, the line central depths, the line position and shifts, etc.

The spectral region containing the interstellar $C_2$ absorption lines of the Phillips system, especially the (3-0) band, is contaminated by the atmospheric lines. Our spectra were divided by a Spica spectrum to remove telluric lines. In this way the majority of telluric lines was removed. The remaining lines were removed individually if they still contaminated our spectra. 
The division of spectra cannot remove all telluric lines. Sometimes very strong atmospheric ones remain but do not blend the measured features. It is appropriate to remove them to make fitting the continuum easier and then measure weak $C_2$ lines. A spectrum is clearer without telluric features. On the other hand, atmospheric lines can also blend measured features and then the latter cannot be taken into consideration. 
The final spectra were normalised to continuum levels to measure the equivalent widths.

The equivalent widths were measured by fitting a Gaussian profile to each $C_2$ line of the Phillips band. In the interstellar medium, because of the very low temperatures and densities, the only efficient mechanism is the Doppler broadening, which allows us to describe the line profiles using a Gaussian function. Then the column density of the rotational level J'' was derived from the equivalent width $W_{\lambda}$\,[m\AA] of the single absorption line using the relationship \citep{Frisch1972}
\begin{equation}
N_{col} = 1.13 \times 10^{17} {\frac {W_{\lambda}}{f_{ij} \lambda^2}} \,,
\end{equation}
where $\lambda$ is the wavelength in [\AA], $f_{ij}$ the absorption oscillator strength.
The band oscillator strengths (Phillips system) were adopted from \citet{Bakker1997}
$f_{10} = 2.38 \times 10^{-3} $, $f_{20} = 1.44 \times 10^{-3}$, 
$f_{30} = 6.67 \times 10^{-4}$. The total column density is the sum derived from the observed J levels.

%
%
\section{Results and discussion}
%
%

\begin{figure}
\centering
\includegraphics[width=0.4\textwidth]{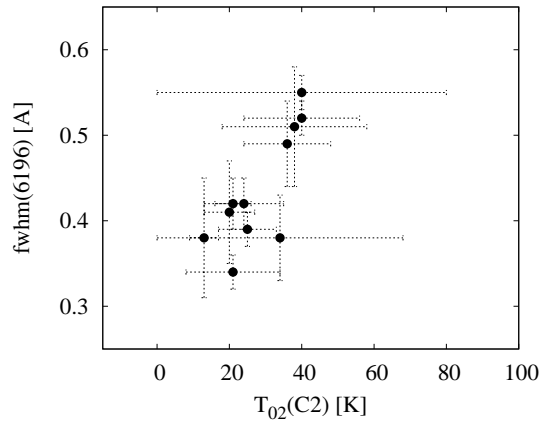}
\caption{A comparison the FWHM of DIB 6196\,\AA~and $T_{02}(C_2)$ toward programme objects.}
\label{fig3}
\end{figure}

We define the rotational temperatures $T_{02}$, $T_{04}$ and $T_{06}$ of the dicarbon molecule as the average excitation temperature of the two, three and four lowest rotational levels of the ground electronic state, respectively. The column densities were used for this purpose. Errors were determined from the standard deviation of the least-square fits.

Generally, rotational temperature defined as above is not the same as the gas kinetic temperature. It is well known from previous work \citep{Dishoeck1982} that populations of all rotational levels cannot be characterised by a single rotational temperature. The lowest J'' levels are described by the lower excitation temperature rather than the higher levels. The model of excitation of the $C_2$ molecule in translucent clouds was presented by \citet{Dishoeck1982} and \citet{Dishoeck1988}. According to this theory, the best approximation of the gas kinetic temperature or its maximum value is provided by the populations of the two lowest rotational levels (J'' = 0 and 2). The higher levels would be populated by the competitive processes of collisional de-excitation and radiative excitation by the average galactic radiation field.

Excitation temperatures ($T_{02}$, $T_{04}$ and $T_{06}$) and column densities were estimated or derived from previous works (see Table 2.). 
For this purpose we used rotational diagrams where \mbox{$-ln[5N_{col}(J'')/(2J''+1)N_{col}(2)]$} is plotted versus E''/k (E'' is the energy of the rotational level J'' and k is the Boltzmann constant) to estimate excitation temperatures \citep{Kazmierczak2009}.

\begin{table*}
\begin{minipage}[]{\columnwidth}
\caption{Summary of observational data}
\label{results}
\begin{tabular}{lccccc}
\hline \hline
object & FWHM(6196)[\AA] & $T_{02}(C_2)^a$[K] &  $T_{04}(C_2)^a$[K] &  $T_{06}(C_2)^a$[K]&  FWHM(CH 4300)[\AA]\\
\hline
HD\,23180 $^{(1}$&$0.41\pm 0.06$&$20 \pm  7$ & $35\pm10$ &$56\pm11$&$0.095\pm0.019$\\
HD\,110432$^{(2}$&$0.38\pm 0.07$&$13 \pm  4$ & $27\pm 6$ &         &$0.120\pm0.014$\\
HD\,147888$^{(3}$&$0.51\pm 0.07$&$38 \pm 20$ & $66\pm13$ &         &$0.101\pm0.013$\\
HD\,147889$^{(4}$&$0.52\pm 0.02$&$40 \pm 16$ & $60\pm 4$ &$69\pm 5$&$0.079\pm0.005$\\
HD\,148184$^{(4}$&$0.55\pm 0.02$&$40 \pm 40$ & $62\pm21$ &$67\pm10$&$0.067\pm0.007$\\
HD\,149757$^{(5}$&$0.49\pm 0.05$&$36 \pm 12$ & $74\pm15$ &         &$0.058\pm0.006$\\
HD\,154368$^{(1}$&$0.42\pm 0.03$&$24 \pm 11$ & $38\pm10$ &$52\pm 6$&$0.076\pm0.006$\\
HD\,163800$^{(4}$&$0.39\pm 0.02$&$25 \pm  8$ & $66\pm21$ &$56\pm 8$&$0.066\pm0.006$\\
HD\,169454$^{(4}$&$0.42\pm 0.03$&$21 \pm  5$ & $29\pm3 $ &$40\pm 3$&$0.082\pm0.007$\\
HD\,179406$^{(4}$&$0.34\pm 0.02$&$21 \pm 13$ & $51\pm13$ &$54\pm 7$&$0.081\pm0.005$\\
HD\,204827$^{(6}$&$0.38\pm 0.05$&$34 \pm 34$ & $42\pm 2$ &$45\pm 2$&$0.146\pm0.015$\\
\hline
\multicolumn{6}{l}{FWHM(6196) and FWHM(CH 4300) were measured in this paper}\\
\multicolumn{6}{l}{$^a$ for $^{(1}$ were measured in this paper}\\
\multicolumn{6}{l}{$^a$ for $^{(2}$ were calculated by Sonnentrucker et al. (2007) using the EW's of van Dishoeck \& Black (1989)}\\
\multicolumn{6}{l}{$^a$ for $^{(3}$ - Sonnentrucker et al (2007)}\\
\multicolumn{6}{l}{$^a$ for $^{(4}$ - Kazmierczak et al. (2009)}\\
\multicolumn{6}{l}{$^a$ for $^{(5}$ were calculated by Sonnentrucker et al. (2007) using the EW's of Hobbs \& Campbell (1982)}\\
\multicolumn{6}{l}{$^a$ for $^{(6}$ were calculated in this work based on EW's from {\'A}d{\'a}mkovics et al. (2003)}\\
\end{tabular}
\end{minipage}
\end{table*}

\begin{table}
\begin{minipage}[]{\columnwidth}
{\vbox{\footnotesize
	\tabcolsep=1.3pt
\centering
\caption{The comparison between our results and those from previous papers.}
\label{comparison}
\begin{tabular}{lllll}
\hline \hline
object & $T_{02} [K]$ & $T_{04} [K]$ & $T_{06} [K]$ &source\\
\hline
HD\,23180 &	       &$48\pm37$ &         &Sonnentrucker et al. (2007)$^a$\\
	  &$20 \pm  7$ &$35\pm10$ &$56\pm11$&this work\\
HD\,147889&$52\pm22$   &$116\pm28$&         &Sonnentrucker et al. (2007)$^b$\\	
          &$40\pm 16$  &$60\pm 9$&$69 \pm 5$&Kazmierczak et al. (2009)\\			
HD\,148184&            &$65$	  &         &van Dishoeck \& de Zeeuw (1984)\\
          &$39\pm16$   &$57\pm12$ &         &Sonnentrucker et al. (2007)$^b$\\	
          &$40\pm 40$  &$62\pm 21$&$67\pm10$&Kazmierczak et al. (2009)\\			
HD\,154368&            &$30$	  &         &van Dishoeck \& de Zeeuw (1984)\\
	  &$27\pm8$    &$38\pm5$  &         &Sonnentrucker et al. (2007)$^b$\\
	  &$24 \pm 11$ &$38\pm10$ &$52\pm 6$&this work\\
HD\,169454&$11\pm3$    &	  &         &Gredel \& Munch (1986)\\	
          &$15^{+10}_{-5}$&	  &         &van Dishoeck \& Black (1989)\\
          &$24\pm6$    &$37\pm5$  &         &Sonnentrucker et al. (2007)$^c$\\	
          &$21\pm  5$  &$29\pm  3$&$40\pm 3$&Kazmierczak et al. (2009)\\			
HD\,179406&	       &$57\pm11$ &         &Sonnentrucker et al. (2007)$^d$\\	
          &$21\pm 13$  &$51\pm 13$&$54\pm 7$&Kazmierczak et al. (2009)\\
HD\,204827&$35\pm  7$  &$51 \pm 3$&	    &Sonnentrucker et al. (2007)\\
	  &$34 \pm 34$ & $42\pm 2$&$45\pm 2$&this work\\		
\hline
\multicolumn{5}{l}{$^a$, $^b$, $^c$, $^d$ calculated by Sonnentrucker et al (2007) using}\\
\multicolumn{5}{l}{the EWs of Hobbs (1981), van Dishoeck \& de Zeeuw (1984),}\\
\multicolumn{5}{l}{Gredel (1999) and Federman et al. (1994) recpectively.}
\end{tabular}
}}
\end{minipage}
\end{table}

\begin{figure*}
\centering
\hbox{
\includegraphics[width=0.49\textwidth]{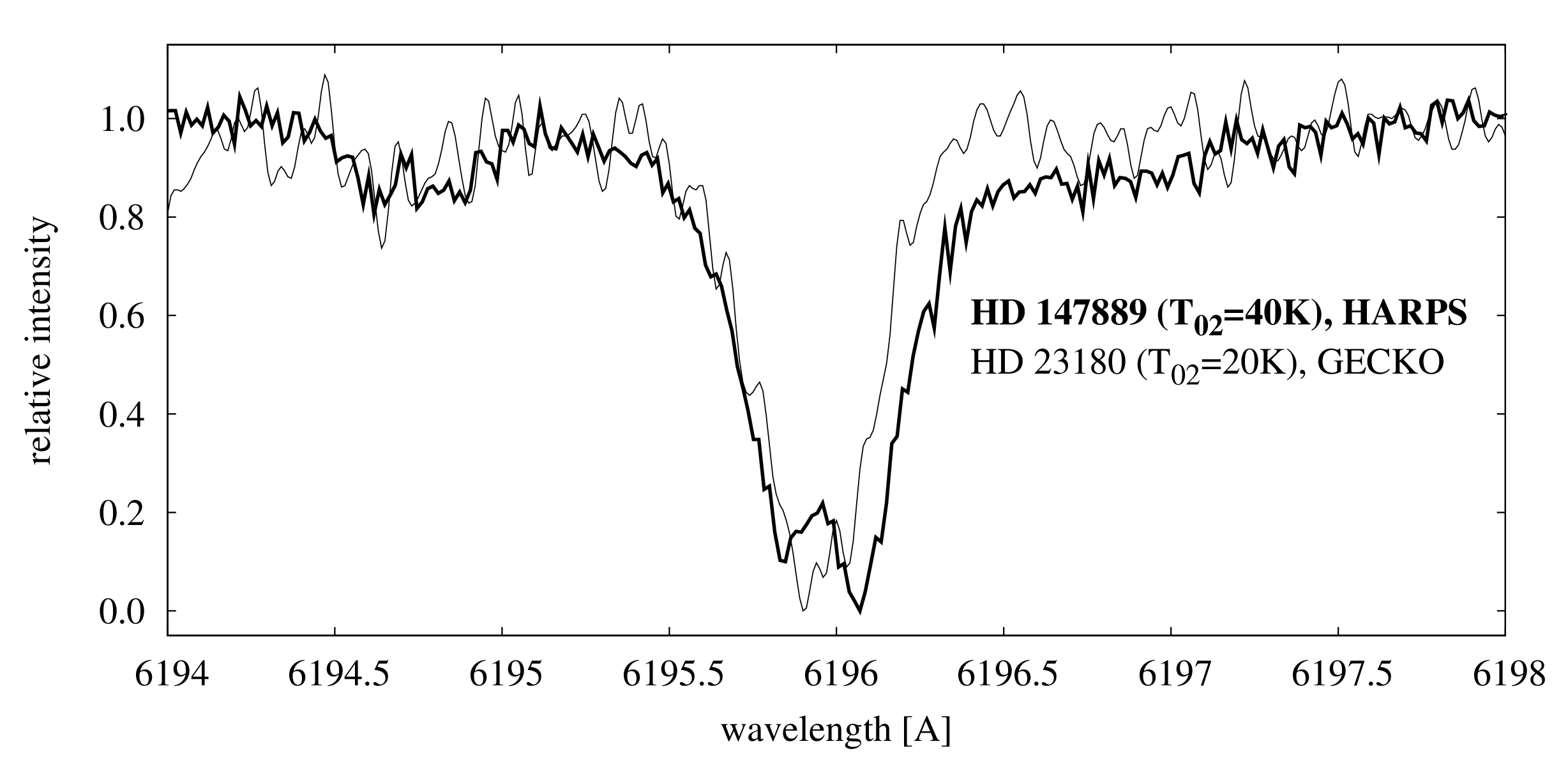}
\includegraphics[width=0.49\textwidth]{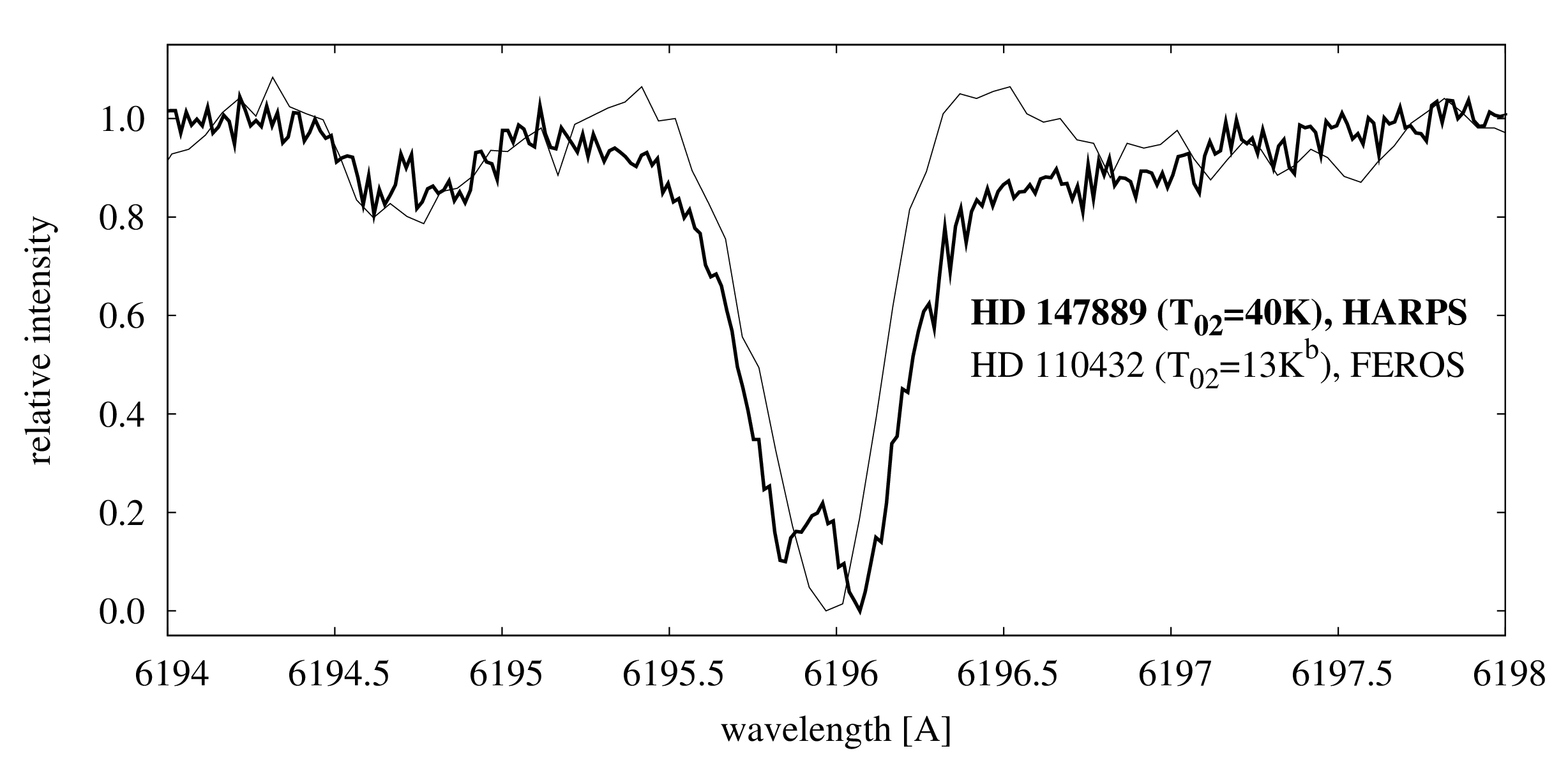}}
\hbox{
\includegraphics[width=0.49\textwidth]{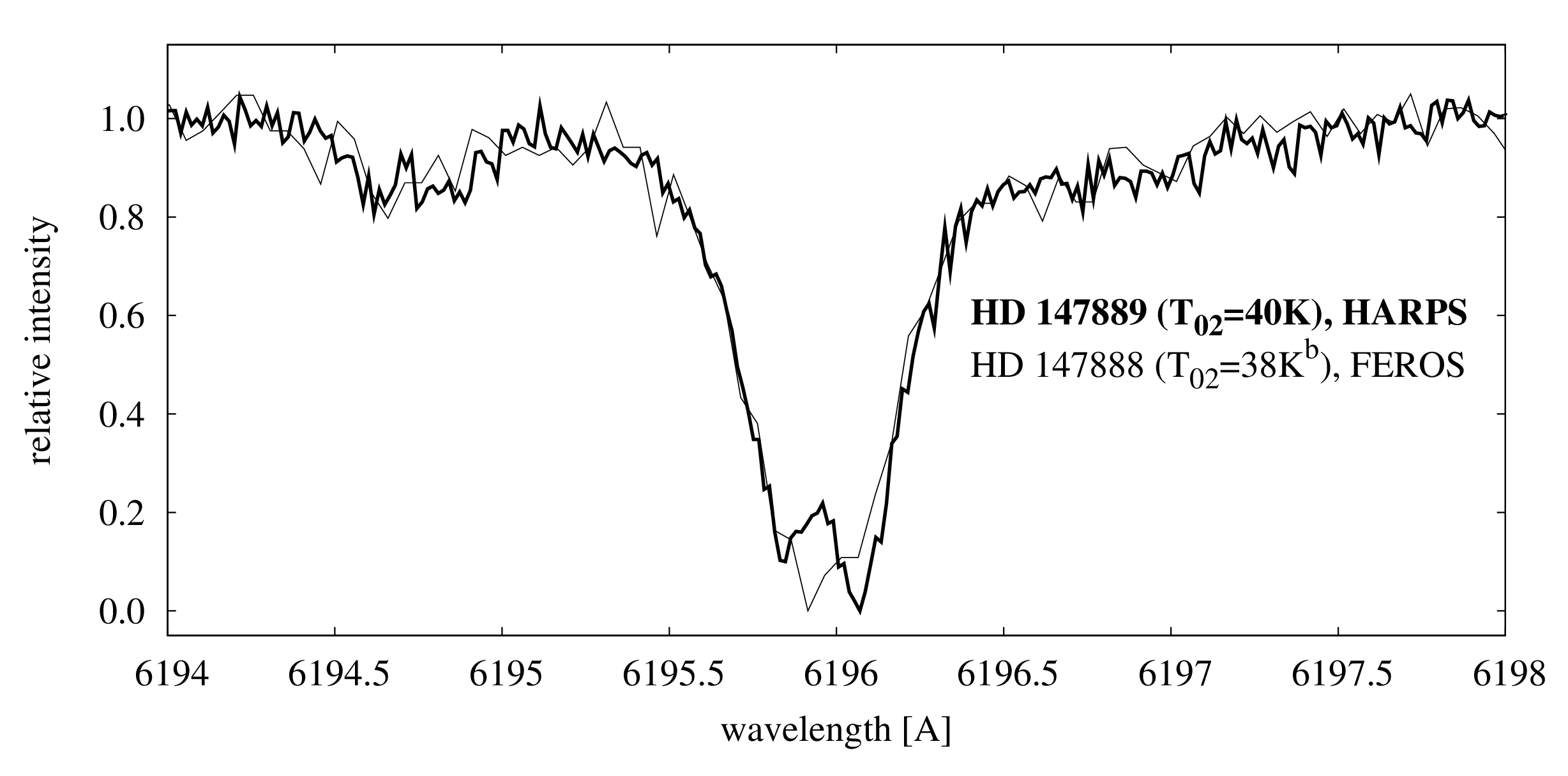}
\includegraphics[width=0.49\textwidth]{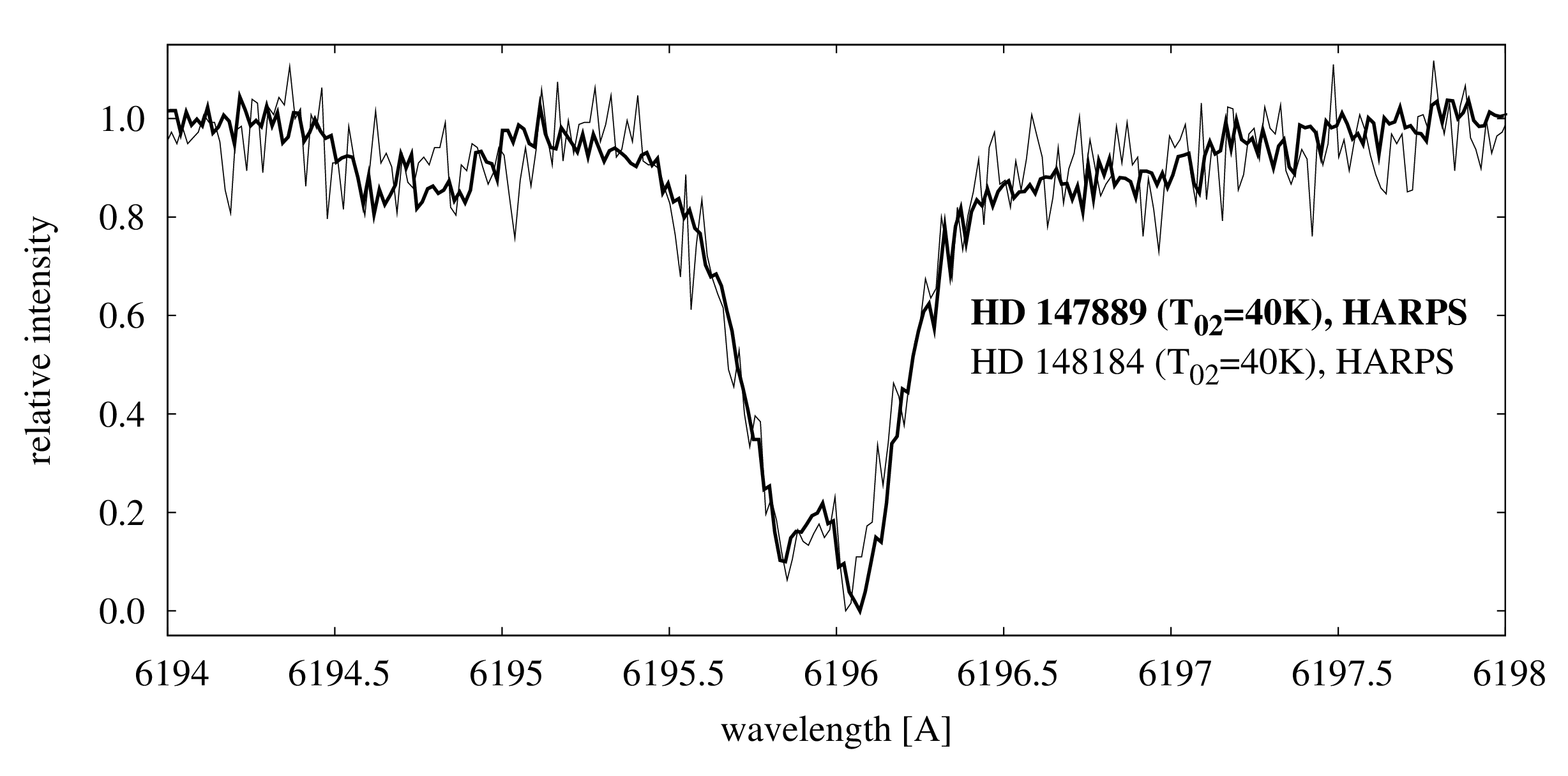}}
\hbox{
\includegraphics[width=0.49\textwidth]{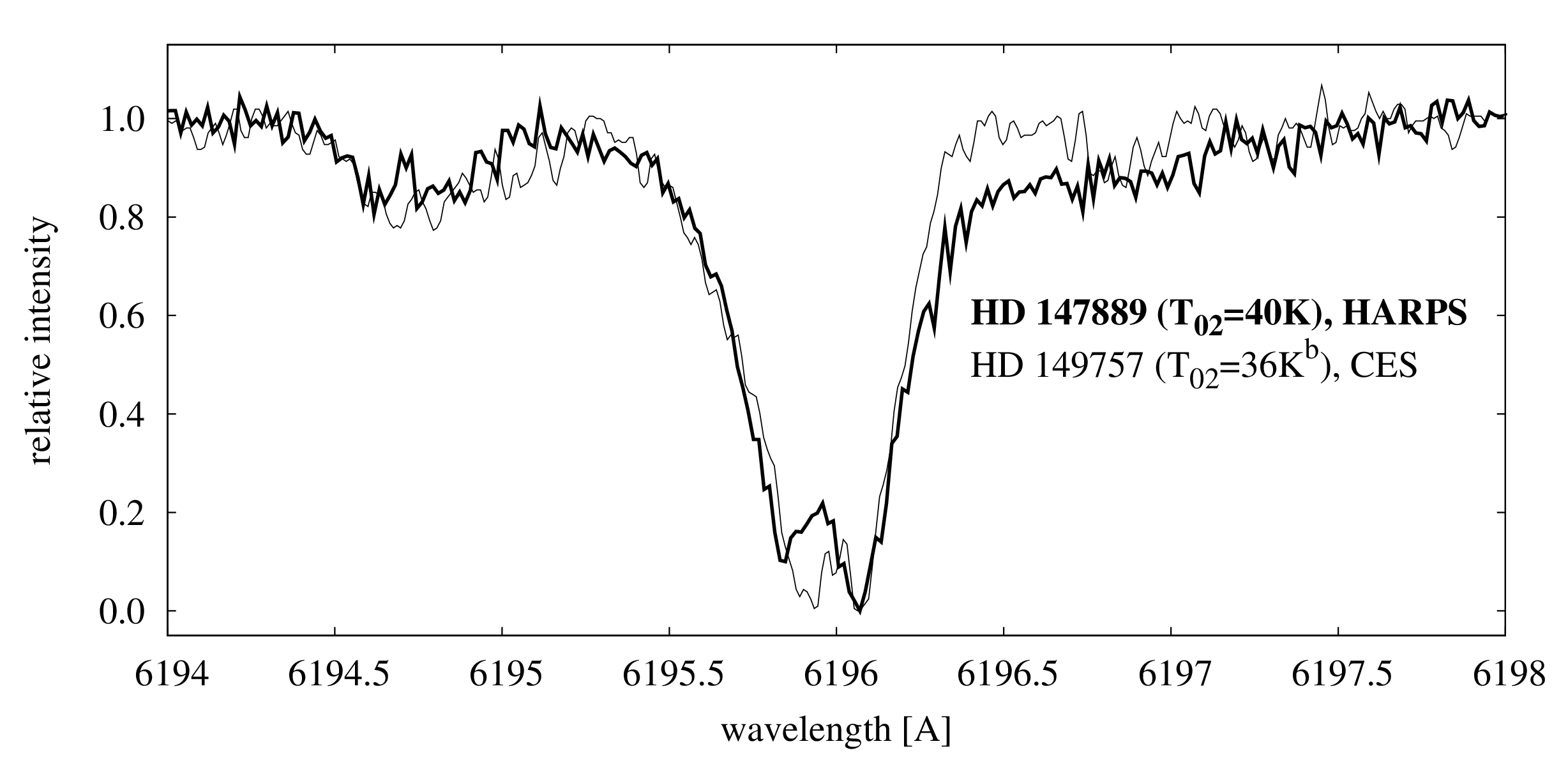}
\includegraphics[width=0.49\textwidth]{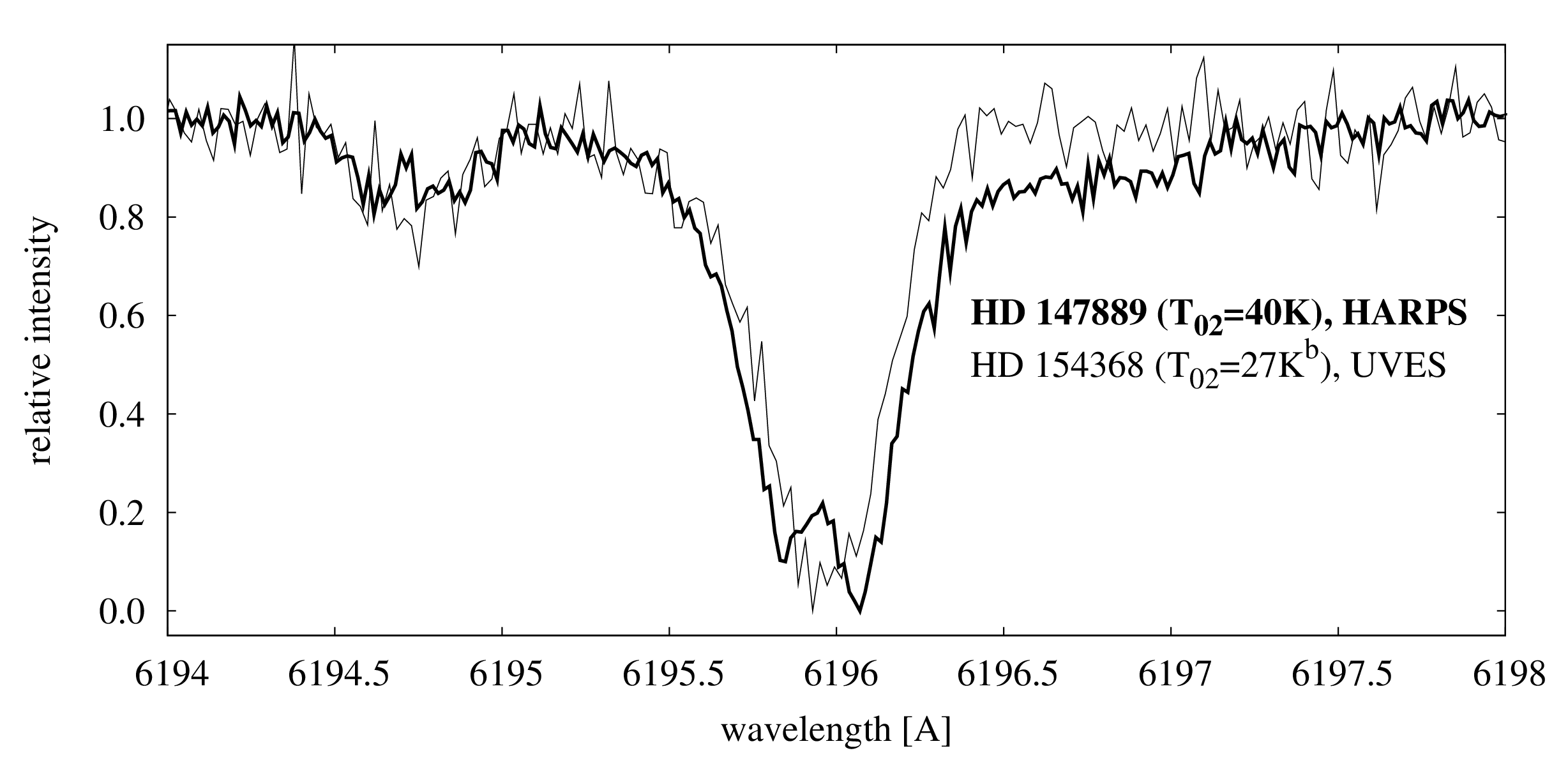}}
\hbox{
\includegraphics[width=0.49\textwidth]{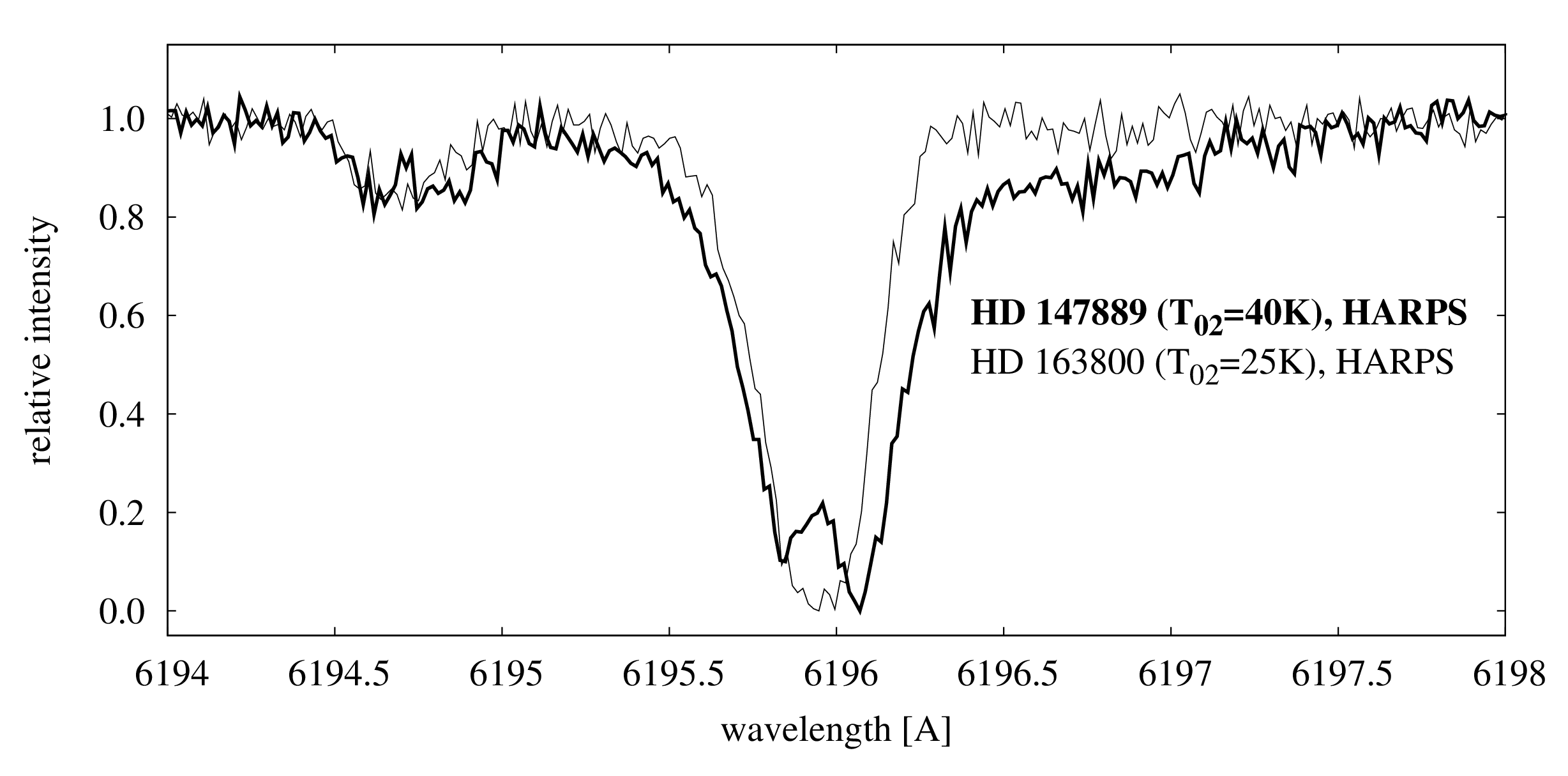}
\includegraphics[width=0.49\textwidth]{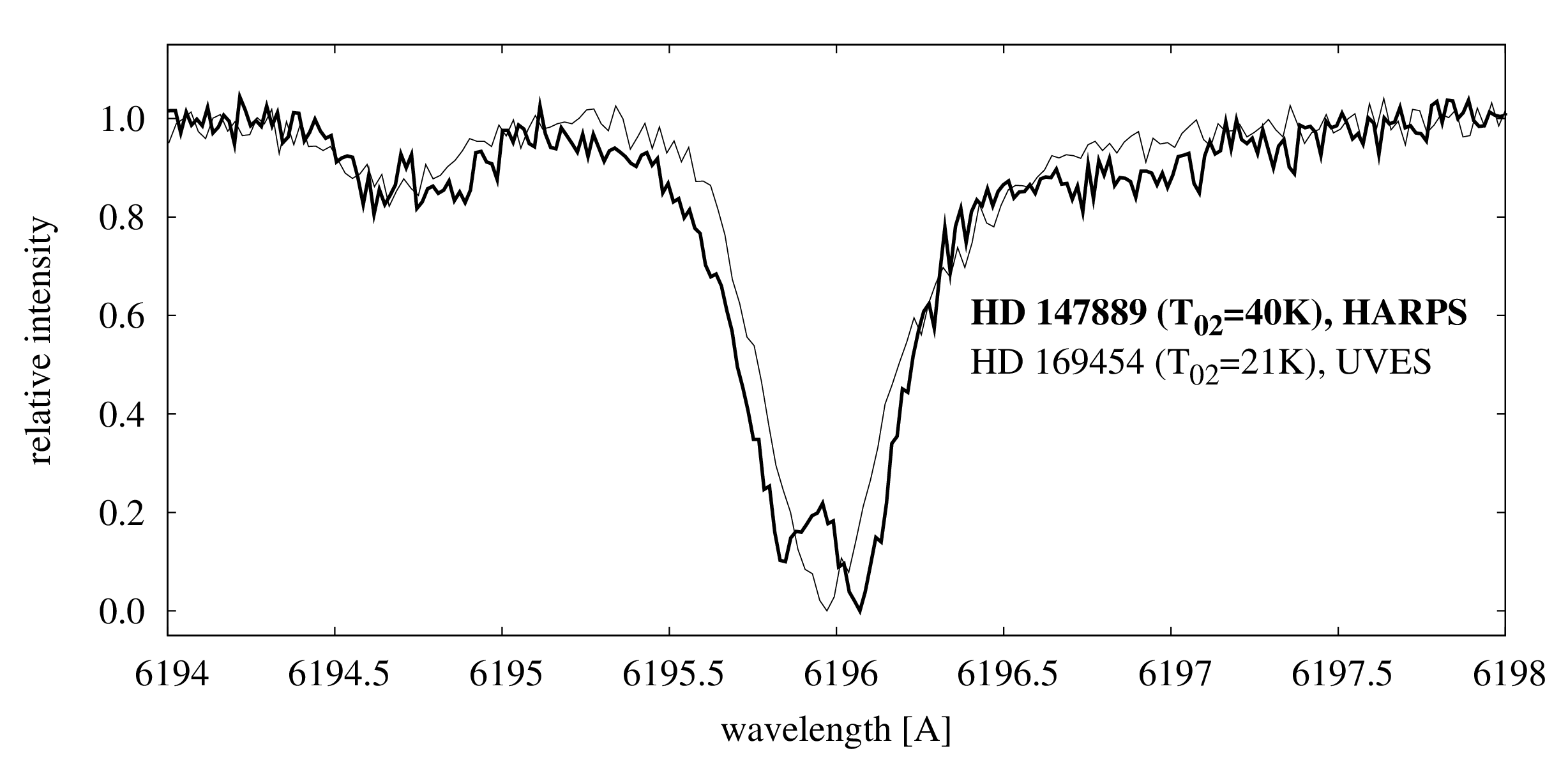}}
\hbox{
\includegraphics[width=0.49\textwidth]{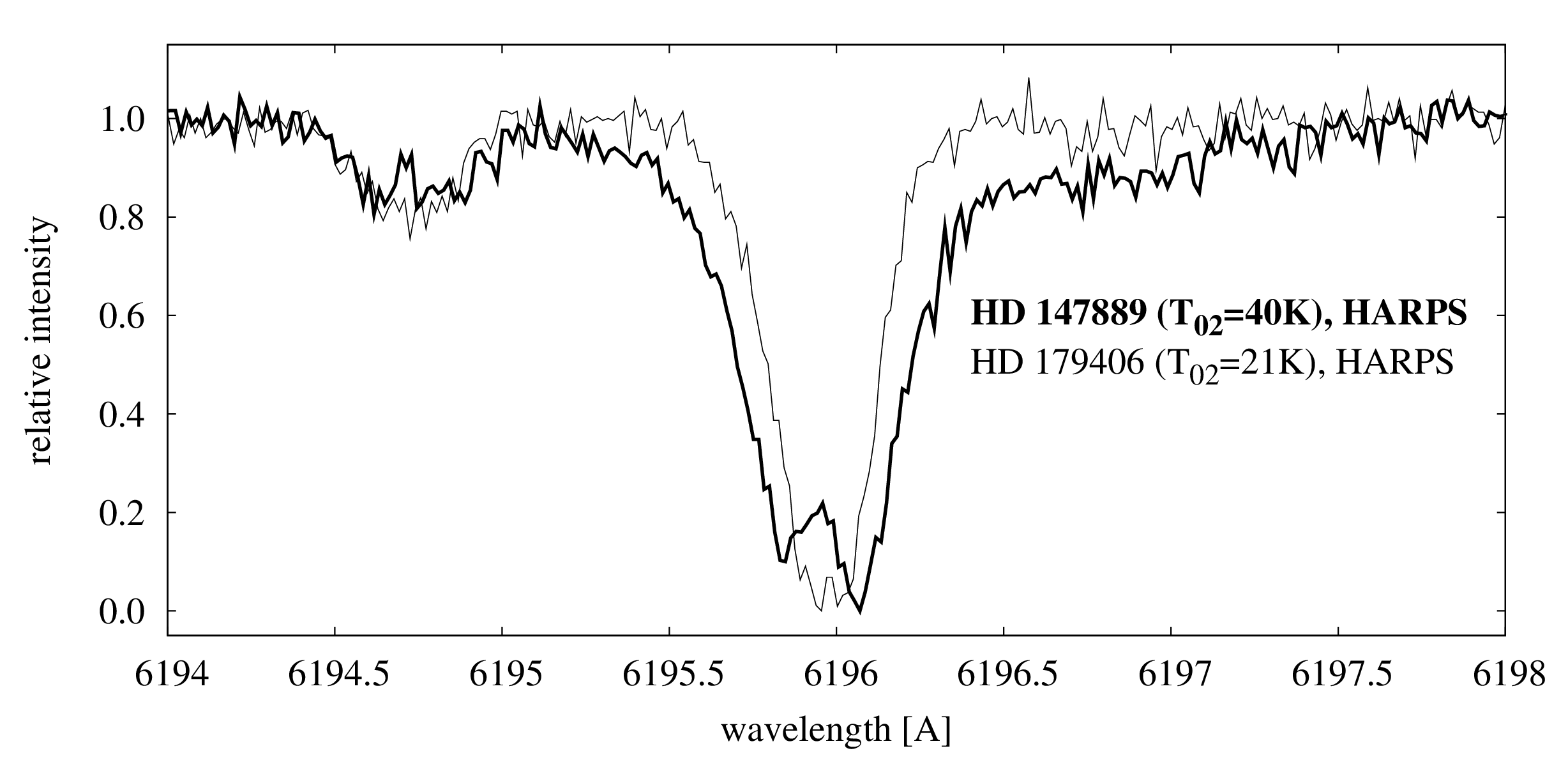}
\includegraphics[width=0.49\textwidth]{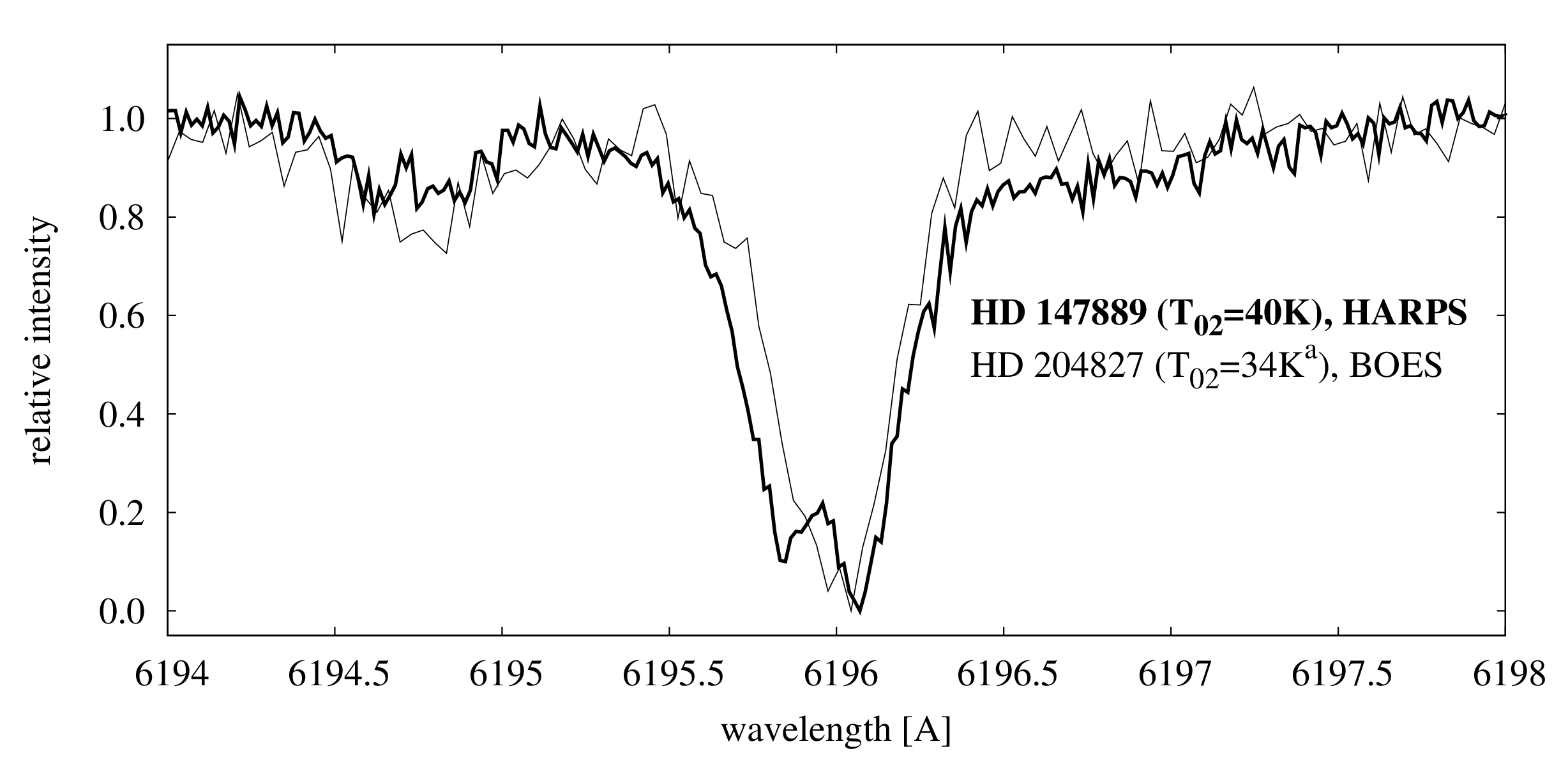}}
\caption{DIB 6196\,\AA~profiles of the spectra, normalised to their central depths, toward programme stars with different rotational temperatures of $C_2$ (* {\'A}d{\'a}mkovics et al. (2003)). DIB profiles depend on the excitation temperature of the dicarbon molecule $T_{02}$. For the higher temperatures they are wider and have some substructures, which disappear for the lower temperatures.}
\label{fig4}
\end{figure*}

To analyse diffuse interstellar bands at 6196\,\AA~we compared their profiles towards program stars and measured the full width at half maximum (FWHM). 

Table 2 presents the results: the full widths at half maximum (FWHM) of DIB at 6196\,\AA\, and the excitation temperatures of $C_2$ ($T_{02}$, $T_{04}$ and $T_{06}$) based on the fit to the first two, three and four rotational levels starting from $J''=0$. Results of $C_2$ in directions to HD\,147889, HD\,148184, HD\,163800, HD\,169454 and HD\,179406 were derived in an earlier work \citep{Kazmierczak2009}; there a detailed comparison was made between the excitation temperature of $C_2$ derived from our data and from other sources; all results are quite similar to values from previous papers (in Table 3, we give the results of $T_{02}$, $T_{04}$ and $T_{06}$) because our results also are based on the excitation temperature. In different publications, these temeperatures are very similar (with one exception). It is difficult to measure temperature from very weak interstellar lines like $C_2$, so the precise value of T is not critical. Much more important is its tendency to be lower or higher. 

In Table 2, we also give the FWHM of CH\,4300\,\AA~to show that there is no correlation between the FWHM(CH 4300) and the excitation temperature of $C_2$ (Fig. 7). In the directions to HD\,147888 and HD\,204827 observations were made with a spectral resolution about 45,000, whereas toward the rest of objects with R=115,000. To remove effects of line broadening from a lower resolving power we measured the FWHM of CH 4300\,\AA~towards HD\,149757 in the spectra with R=115,000 and 45,000; the relation of the FWHM for different resolutions is 1.76; we divided by this value by the full widths at half maximum of CH 4300\,\AA~in the directions of HD\,147888 and HD\,204827 before being plotted (Figure 7).

The main result is shown in Figure 3 (and in detail in Fig. 4). The correlation between the full width at half maximum of the DIB at 6196\,\AA~and the excitation temperature of $C_2$ ($T_{02}$) is presented. A similar relation was drawn for $T_{04}$ (Fig. 5) and $T_{06}$ (Fig. 6). The best correlation is shown for $T_{02}$, but the others are also quite good. 

\begin{figure}
\centering
\includegraphics[width=0.36\textwidth]{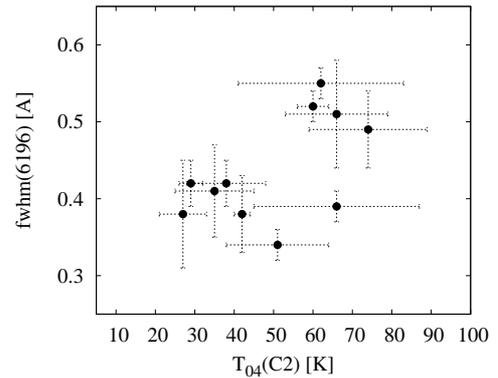}
\caption{The profile width of the DIB 6196\,\AA~varie's from object to object, likely because of the varying excitation temperature $T_{04}$ of $C_2$, it being wider for higher temperatures.}
\label{fig5}
\end{figure}

\begin{figure}
\centering
\includegraphics[width=0.36\textwidth]{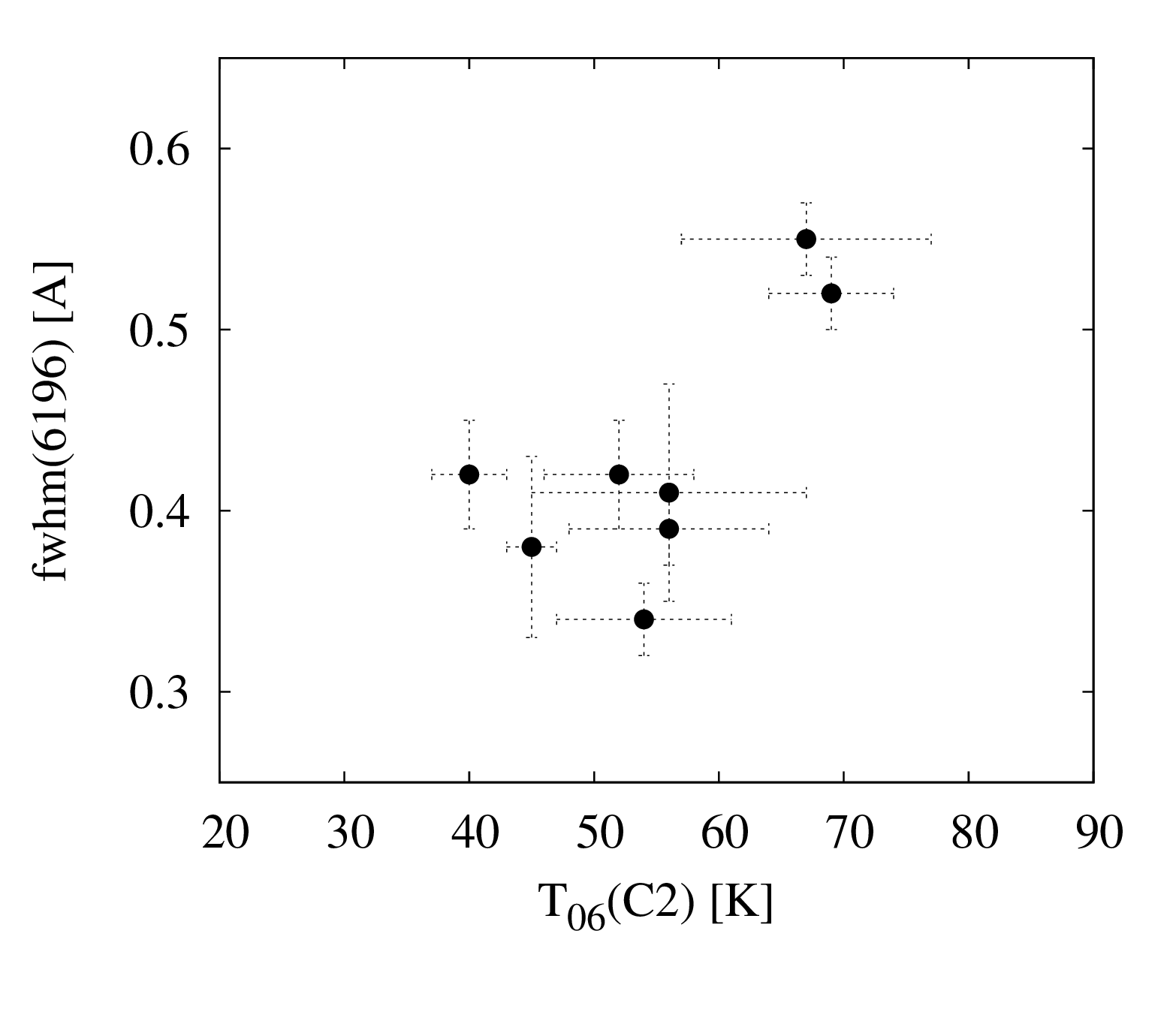}
\caption{The profile width of the DIB 6196\,\AA~varie's from object to object, likely because of the varying excitation temperature $T_{06}$ of $C_2$, it being wider for higher temperatures.}
\label{fig6}
\end{figure}

\begin{figure}
\centering
\includegraphics[width=0.36\textwidth]{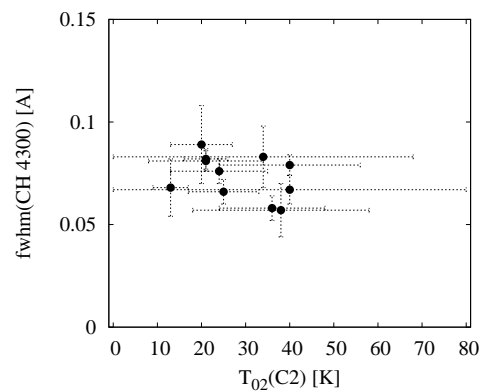}
\caption{A comparison of the FWHM of CH 4300\,\AA~line and the excitation temperature of $C_2$ toward the programme objects. There is no correlation.}
\label{fig7}
\end{figure}

\begin{figure}
\centering
\includegraphics[width=0.4\textwidth]{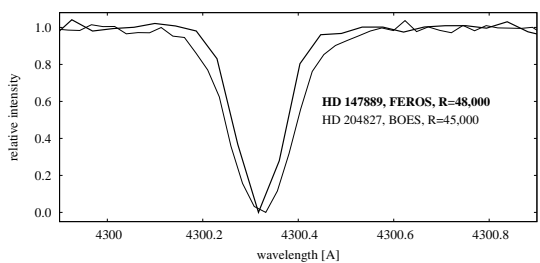}
\caption{A comparison of the width profile of the CH 4300\,\AA~ feature toward two stars (HD\,147889 and HD\,204827) with almost the same spectral resolution.}
\label{fig8}
\end{figure}

%
%
\section{Conclusions}
%
%

The high resolution and high signal-to-noise ratio spectra acquired using different instruments leave no doubt that the profile width of the DIB at 6196\,\AA~varie's from object to object.

It was suggested by \citet{Galazutdinov2002} that this band is very narrow and almost feature-less towards some objects (e.g. HD\,179406) while toward others (e.g. HD\,149757) it is evidently broader and splits into some substructures at the bottom of the line. There are at least two mechanisms which could explain the variations of DIB profiles. One of these is the shift and intensity variation of the isotopic components, but that depends on the relative content of $^{13}C$ atoms in the interstellar molecules, and it leaves the width of profile unchanged \citep{Webster1996}. The second is the variation of the rotational temperature of the carrier species \citep{Cami2004}; this variability could be related to the excitation temperature of the dicarbon molecule or other simple centrosymmetric species.

Figure 4 compares the 6196\,\AA~region in the spectra towards our target stars. One important conclusion immediately drawn is that the profiles of the DIBs at 6196\,\AA~depend on the excitation temperature of $C_2$. In the case where this temperature is very similar (toward HD 147889 and HD 148184) the profiles of the DIB at 6196\,\AA~are of almost the same width. 
At higher temperatures the profiles of 6196 are wider and contain some substructures \citep{Galazutdinov2002} which disappear in the case of lower temperatures. 
The 6196 wavelength shifts seen in Figure 4 are most likely due to different excitation of the DIB carrier. The effect cannot be precisely interpreted until the carrier is identified. The shifts are not caused by the Doppler effect; they are not observed while the profiles are identical. Figures 3 and 7 clearly show that the broadening of 6196 is not a result of Doppler splitting since it is absent in the much narrower $CH$ features. The $CH$ line was shown to be Doppler split in HD\,204827 \citep{Pan2004}; in our spectra it is broader than in HD\,147889 (Fig. 8) while the 6196 feature is narrower (Fig. 4). Figure 1 also shows that there are no widely separated Doppler components but unfortunately there still could be multiple components at the same velocity and features originating from different physical conditions could be blended. This possible phenomenon likely creates the scatter observed in Figures 3, 5, 6. A mixture of high and low excitation temperature spectra should lead to an excitation temperature higher than the lowest temperature case which likely shrinks the T range in Fig. 4.

Our results may suggest that carriers of some DIBs, especially 6196\,\AA, could be centrosymmetric molecules, whose spectral features become broader as their rotational temperatures increase. This means that conditions of excitation of $C_2$ and the DIB 6196\,\AA~carrier should be similar.

Moreover our results (e.g. temperatures) for $C_2$ are consistent with those for $C_3$ \citep{Adamkovics2003}. The profile of the narrow DIB at 6196\,\AA~apparently depends on the rotational temperature estimated from the dicarbon molecule, being broader for higher temperatures, which is characteristic of both $C_2$ and $C_3$.
 
$C_3$ was analysed by \citet{Adamkovics2003} and \citet{Oka2003}. They found that the column density and rotational temperature derived from $C_3$ is well correlated with the same parameters of $C_2$.
It is interesting and important to analyse the simplest multicarbon chains (like $C_2$ or $C_3$), because they give information about physical conditions in interstellar clouds. 
The observed profile variations of the 6196 DIB rule out polar molecules as a possible carrier. Such species are efficiently cooled by rotational emission and profiles of their bands should be nearly the same in all objects.

A broader survey of other DIBs in as large a sample as possible is highly desirable. Unfortunately it is  difficult to find objects with little Doppler splitting in the interstellar lines; only in this case is a comparison of physical parameters  possible. Furthemore, a selection of single cloud cases is possible only after very high resolution observations of many targets are made.

%
%
\begin{acknowledgements}
%
%
This work was supported by the Science and High Education Ministry of Poland, 
grants N203 012 32/1550, N203 019 31/2874 and N203 39/3334.
\end{acknowledgements}

%
%
\bibliographystyle{aa}
\bibliography{1558}
\end{document}